\documentclass[fleqn,10pt]{wlscirep}

\usepackage[utf8]{inputenc}
\usepackage[T1]{fontenc}
\usepackage[english]{babel}
\usepackage{marvosym}
\usepackage{amsfonts,bm}
\usepackage{amsmath}
\usepackage{amssymb}
\usepackage{graphicx}
\usepackage{fancyhdr}
\usepackage{float}
\usepackage{braket}
\usepackage[caption=false]{subfig}
\usepackage{tikz}  
\usepackage{chngcntr}
\usepackage{ulem}

\usepackage{color}

\usepackage{hyperref}

\def \ufro {Departamento de Ciencias Físicas, Universidad de La Frontera, Casilla 54-D, 4811186 Temuco, Chile.}
\def \cedenna {Center for the Development of Nanoscience and Nanotechnology (CEDENNA), 9170124 Santiago, Chile}
\def \ucen {Escuela de Ingenier\'ia, Universidad Central de Chile, Avda. Santa Isabel 1186, 8330601 Santiago, Chile}
\def \bari {Department of Electrical and Information Engineering, Technical University of Bari, Bari, 70125, Italy.}
\def \messina {Department of Engineering, University of Messina, 98166, Italy.}
\def \messinab {Department of Mathematical and Computer Sciences, Physical Sciences and Earth Sciences, University of Messina, Messina, I-98166, Italy.}
\def \delft{Delft University of Technology, Delft, The Netherlands}
\def \usm{Universidad Técnica Federico Santa María, Avenida España 1680, 2390123 Valparaiso, Chile}

\begin{document}
\title{Modeling the spatial resolution of magnetic solitons in Magnetic Force Microscopy and the effect on their sizes}
    \author[1]{I. Castro}

    \author[2,$\dagger$]{A. Riveros}

   \author[2,3]{J. L. Palma}

    \author[4]{L. Abelmann}
  
    \author[5]{R. Tomasello}
    \author[5]{D. R. Rodrigues}
  \author[6]{A. Giordano}
    \author[7]{G. Finocchio}
  \author[8]{R. Gallardo}
    \author[1,*]{N. Vidal-Silva}

\affil[1]{\ufro}
\affil[2]{\ucen}
\affil[3]{\cedenna}
\affil[4]{\delft}
\affil[5]{\bari}
\affil[6]{\messina}
\affil[7]{\messinab}
\affil[8]{\usm}
\affil[*]{nicolas.vidal@ufrontera.cl}
\affil[$\dagger$]{alejandro.riveros@ucentral.cl }

\begin{abstract}
In this work, we explored theoretically the spatial resolution of magnetic solitons and the variations of their sizes when subjected to a Magnetic Force Microscopy (MFM) measurement. Next to tip-sample separation, we considered reversal in the magnetization direction of the tip, showing that the magnetic soliton size measurement can be strongly affected by the magnetization direction of the tip. In addition to previous studies that only consider thermal fluctuations, we developed a theoretical method to obtain the minimum observable length of a magnetic soliton and its length variation due to the influence of the MFM tip by minimizing the soliton's magnetic energy.  Our model uses analytical and numerical calculations and prevents overestimating the characteristic length scales from MFM images. We compared our method with available data from MFM measurements of domain wall widths, and we performed micromagnetic simulations of a skyrmion-tip system, finding a good agreement for both attractive and repulsive domain wall profile signals and for the skyrmion diameter in the presence of the magnetic tip. Our results provide significant insights for a better interpretation of MFM measurements of different magnetic solitons and will be helpful in the design of potential reading devices based on magnetic solitons as information carriers.
\end{abstract}

\maketitle


\section*{Introduction}
The goal of miniaturizing magnetic elements in efficient micro- and nanodevices is driving advances in synthesis and characterization techniques. Magnetic force microscopy (MFM) \cite{kazakova2019frontiers,koblischka2003recent,vokoun2022magnetic,Abelmann2017,abelmann2005towards} stands out as a versatile and straightforward method for imaging magnetic textures within samples, as evidenced by various studies \cite{manke2010three,szmaja2006recent,birch2020real,boulle2016room,meng2019observation,okuno2002mfm}. Its working principle is based on the magnetic interaction between a small magnetic element (the tip) mounted on a cantilever spring, and a magnetic sample, which should be thin enough to consider no significant variations of the sample's magnetization within the volume. The physics behind this interaction relies on the magnetic stray field generated by the magnetic sample, which is detected by the tip through the magnetostatic interaction energy \cite{rugar1990magnetic,schonenberger1990understanding,porthun1998magnetic,hartmann1999magnetic}. How this detection takes place corresponds to variations in the cantilever oscillation frequency due to force gradients on the magnetic tip generated by the vertical displacement of the cantilever's end.
The magnetic tip should also be close enough to get a clear image but not so close to prevent deformations on the sample magnetization \cite{abelmann2005towards}. By assuming that variations in tip height are small compared to the tip-sample distance, the problem of theoretically characterizing an MFM measurement reduces to calculating the magnetic force exerted over the sample or the tip. There are several reports in which the magnetic force is obtained for different geometrical and magnetic configurations between the tip and the sample by using distinct approximations \cite{Abelmann2017,cambel2010magnetic,kuramochi2007advantages,porthun1998magnetic,porthun1998optimization,saito2003high,tanaka2012theoretical,wadas1992models,wolny2010iron}. In most of them, the calculation of the magnetostatic interaction energy between tip and sample is carried out on the tip volume, which requires knowing the tip's magnetization and the stray field generated by the sample. The typical assumption is modeling the sample's magnetization pattern as a periodically variable one \cite{abelmann2005towards}, which facilitates the calculation of the stray field when transforming into the Fourier space. These assumptions indeed allow defining the so-called tip transfer function (TTF) \cite{hug1998quantitative}, a quantity that relates the magnetic force (in the frequency domain) with the sample's magnetization. By using the \textit{monopole} approximation \cite{Abelmann2017,hartmann1999magnetic,wolny2010iron,grutter1992magnetic,van2000method}, the TTF can be readily calculated providing an estimate of the MFM sensitivity can be obtained \cite{albrecht1991frequency} for samples with periodic magnetic patterns.\\
However, when considering localized magnetic patterns such as magnetic solitons, calculations in Fourier space are less efficient and the theoretical analysis becomes more convenient to be performed in real space.  A magnetic soliton is a localized magnetic texture characterized by a typical length scale. \cite{guimaraes2009principles,aharoni2000introduction,wang2018theory}. Examples of magnetic solitons are domain walls \cite{parkin2008magnetic,guimaraes2009principles} and magnetic skyrmions  \cite{fert2017magnetic,nagaosa2013topological}, whose characteristic lengths are the domain wall width and the skyrmion diameter, respectively. Besides, magnetic solitons are keys for nanodevices such as nanosensors, nanotransistors, and nano oscillators, among others \cite{yu2021magnetic,fert2017magnetic,saavedra2021magnonic,zhang2015skyrmion,siddiqui2019magnetic}. An MFM measurement of such magnetic configurations should be performed carefully, as the characteristic length could suffer variations produced by the interaction with the tip. These variations are significant compared to the ones caused by thermal fluctuations. Indeed, the latter are negligible since, otherwise, MFM measurements would not be possible. In this frame, the monopole approximation and the use of the TTF (obtained from a periodic magnetization pattern) for calculating the MFM sensitivity breaks down. Therefore, a theoretical analysis that considers the tip's influence on the texture characteristic length and enables the determination of MFM sensitivity is crucial for experiments involving magnetic solitons.\\

 In this manuscript, we analytically investigate both the sensitivity of MFM in spatial resolution and the impact of MFM measurements on the size of magnetic solitons, such as DWs and skyrmions. We compare our results with previous experimental studies on MFM regarding DW stripes and with micromagnetic simulations of the skyrmion-tip system. Our results show strong agreement with both simulation and experimental data.

\section*{Model}
To study analytically and numerically both the sensitivity of MFM in spatial resolution and the effect of an MFM measurement on the sample's magnetization, we consider the interaction energy between the tip and sample. Due the reciprocity principle \cite{aharoni2000introduction}, the magnetostatic interaction energy can be calculated by two ways: by integrating the coupling between the sample dipolar field  and tip magnetization in the tip volume space, or by integrating the coupling between the tip dipolar field and sample magnetization in the sample volume space, i.e., $\label{eintgeneral}
    E_{\text{int}}= -\mu_0\int_{\text{tip}}\mathbf{M}_{\text{t}}\cdot\mathbf{H}_{\text{s}} dV_{\text{t}}= -\mu_0\int_{\text{sample}}\mathbf{M}_{\text{s}}\cdot\mathbf{H}_{\text{t}} dV_{{\text{s}}}$, where $\mathbf{M}$ and $\mathbf{H}$ are, respectively, the magnetization field and the generated stray field. The sub indices `s' and `t' stand for \textit{sample} and \textit{tip}, respectively. The magnetic force is then obtained by applying  gradient, $\mathbf{F} = -\mathbf{\nabla}E_{{\text{int}}}$. The MFM sensitivity can be obtained by comparing the vertical force component over tip due to magnetostatic interaction with thermal effects \cite{albrecht1991frequency,abelmann2005towards}, giving a minimum detectable MFM signal, $l_{\text{th}}$. Nevertheless, according to the Newton's third law, this can be calculated by the vertical force component over the sample ($F_{s_z}$) such that $l_{\text{th}}$ can be obtained through:
\begin{align}
    \biggl( \frac{\partial F_{\text{s}_z}}{\partial d_z} \biggl)^2 = \frac{4k k_\text{B} T B}{\omega_0 Q A^2},
    \label{eq:thermal_noise}
\end{align}
 where $k$ is the cantilever spring constant, $T$ is the temperature at which the measurement is carried out, $B$ the bandwidth related to the thermal noise source, $\omega_0$ is the cantilever resonance frequency, $Q$ the quality factor, $A$ the mean square amplitude, and $k_\text{B}$ is the Boltzmann constant.
\begin{figure}[h!]
\centering
\includegraphics[width=14cm]{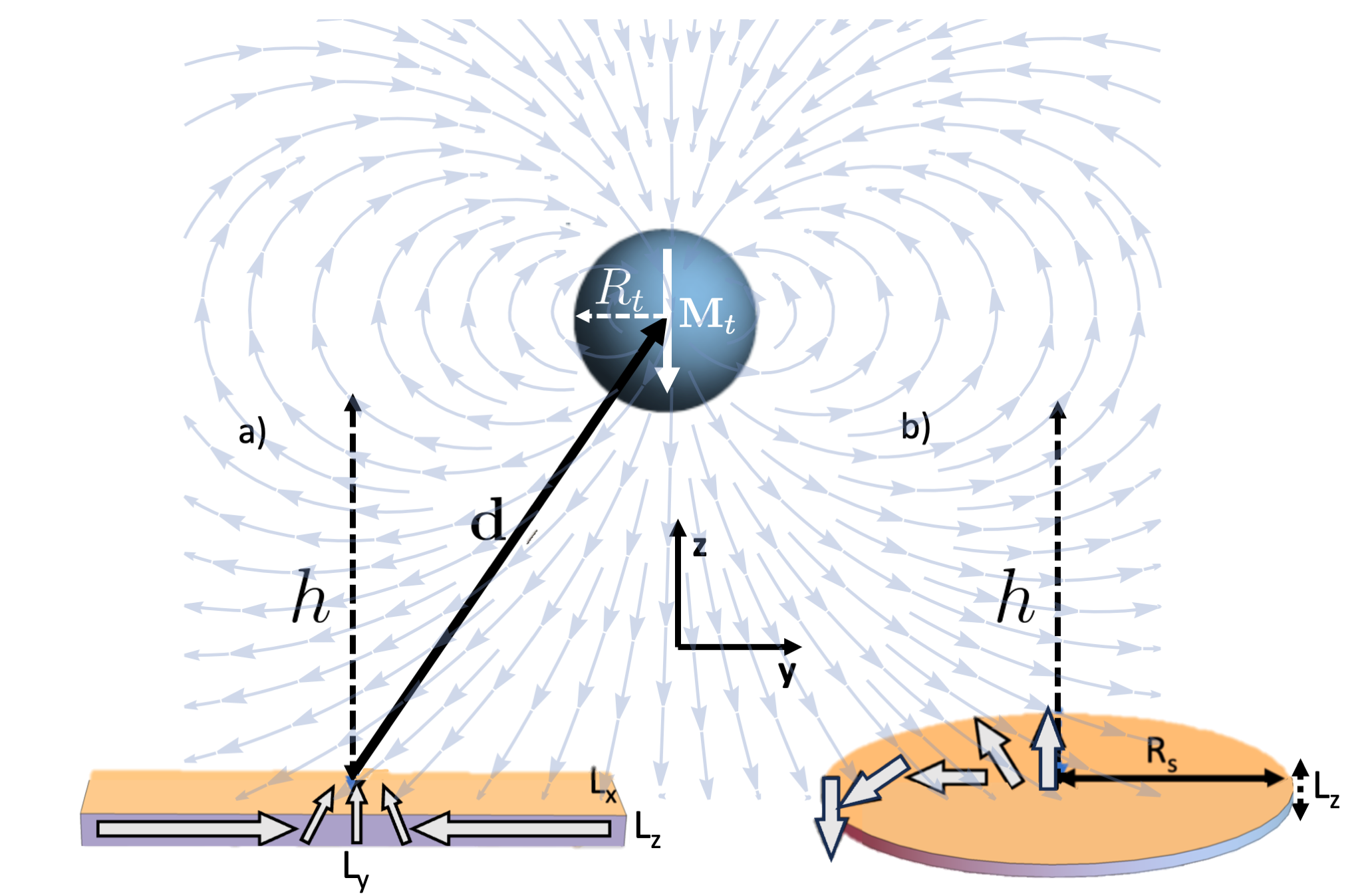}
\caption{Schematic of a spherical MFM-like tip of radius $R_\text{t}$ and $\mathbf{M}_\text{t}=-M_\text{t}\hat{k}$ acting on a) a rectangular thin stripe with dimensions $L_x,L_y,L_z$ and b) a thin nanodisk with radius $R_\text{s}$ and thickness $L_\text{z}$. In both cases, the distance of separation is $h = d_z-R_\text{t}$, and vector $\mathbf{d}$ is measured from the center of the magnetic texture to the tip center. The arrows in the background represent the stray field generated by the spherical tip, while the gray ones within the both samples depict their magnetization fields (DW and skyrmion), respectively.}
\label{fig:schematic}
\end{figure}
The method to obtain the sensitivity in spatial resolution lies in the heart of Eq. \eqref{eq:thermal_noise}, together with the implementation of any Ritz models to describe the magnetization of the sample, which is defined as a field vector that depends on some Ritz parameters $ l_{\text{ch}}$ corresponding to minimizable parameters that characterize the size of the magnetic texture hosted in the sample. Therefore, for a given geometry and magnetic parameters of the sample and tip, the left hand side of Eq. \eqref{eq:thermal_noise} will be a function of $ l_{\text{ch}}$ and the distance $\mathbf{d}$  between the tip and sample (see Fig. \ref{fig:schematic} and supplementary material). By solving numerically Eq. \eqref{eq:thermal_noise}, $l_{\text{ch}}$ will correspond to the minimum detectable length, $l_{\text{th}}$, that the thermal noise allows to measure by MFM for a given cantilever parameters and temperature.
Besides, the method also allows the analysis of the effects of a MFM measurement on the magnetization of the sample by minimizing the total magnetic energy of the sample, $E_{\text{s}}=E_\text{self}+ E_\text{int}$, as a function of the Ritz parameters $ l_{\text{ch}}$. The term $E_{\text{self}}$ may consist of exchange, anisotropy, dipolar,  Dzyaloshinskii-Moriya Interaction (DMI), and Zeeman energies. The interaction energy between the tip and sample is given by $E_{\text{int}}$. By denoting $l_{\text{ph}}$ as the physical value of the Ritz parameters that minimizes $E_{\text{s}}$, then we can compare the minimum detectable length allowed by thermal noise, $l_{\text{th}}$, with the corresponding physical length, $l_{\text{ph}}$. To get accurate measurements, at least $l_{\text{ph}}$ should be equal or greater than $l_{\text{th}}$, i.e., $l_{\text{th}}\leq l_{\text{ph}}$. This is an important point, giving some insights into the magnetic texture's possible deformation due to the magnetic tip's presence in a typical MFM measurement.

\section*{Results and Analysis}
 We apply the model described above to characterize an MFM-like measurement of a domain wall in thin magnetic stripes and skyrmions in thin multilayered circular dots with out-of-plane magnetic anisotropy. The geometrical parameters for the sample and tip are depicted in Fig. \ref{fig:schematic}. Here, we model the tip as a spherical tip of radius $R_\text{t}$, uniformly magnetized along the $\pm z-$direction, i.e.,  $\mathbf{M}_\text{t} = \pm M_\text{t}\hat{k}$, being $M_\text{t}$ the saturation magnetization of the tip. Note that we model the magnetic tip as a dipole.
For these systems, the $z-$ component of the magnetic force applied over the sample can be written as:
\begin{align}
F_{\text{s}_z} = \pm \mu_0 M_\text{s} M_\text{t} \int_\text{sample}  \hspace{-0.2cm}\mathbf{m}_\text{s}(\mathbf{r},l_\text{ch})\cdot \partial_z \mathbf{h}_\text{t}(\mathbf{r}-\mathbf{d})\, dV_\text{s}
\label{eq_Fz_sample_both_textures}
\end{align}
 respect to a reference system fixed at the sample (see Supplementary material). In Eq. \eqref{eq_Fz_sample_both_textures}, $\mathbf{m}_\text{s}=\mathbf{m}_\text{s}/M_\text{s}$ is the normalized sample magnetization and $\partial_z\equiv\partial/\partial z$. Besides, we have defined the dimensionless stray field generated by the tip as $\mathbf{h}_\text{t}=\mathbf{h}_\text{t}/M_\text{t}$, and $\pm$ signs hold for tip magnetization in $\pm z$-direction.
We use a spherical tip because the ends of various tip geometries can be modeled as magnetic spheres due to their different curvature radii. Additionally, the calculation is greatly simplified by considering this spherical geometry, which allows us to deal with closed analytical expressions for $\mathbf{h}_\text{t}$, keeping the general effects unalterable. The stray field generated by the tip outside the sphere can be found in the Supplementary Information.
\begin{figure}[h!]
\centering
\includegraphics[width=13cm]{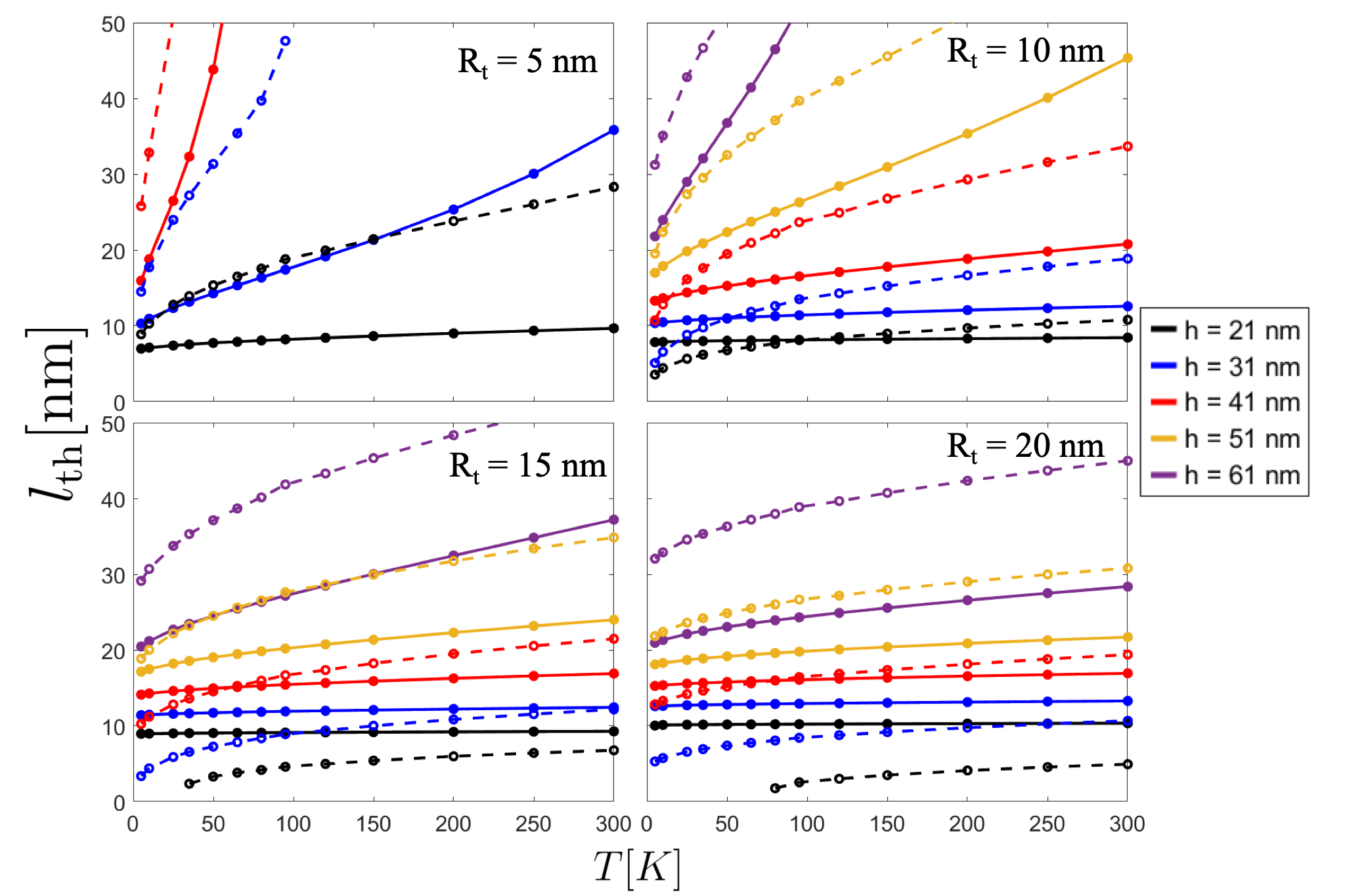}
\caption{Minimum detectable MFM signal allowed by the thermal noise, $l_{\text{th}}$, as a function of temperature for different distances of separation and a) $R_\text{t} = 5$ nm, b) $R_\text{t} = 10$ nm, c) $R_\text{t} = 15$ nm and d) $R_\text{t} = 20$ nm. Filled circles with full lines correspond to DW, while open circles with dashed lines hold for skyrmion sample.}
\label{fig:lth_T}
\end{figure}
In the following, we analyze the sensitivity in spatial resolution and the effect of MFM measurements over DWs and Skyrmions. The details of the sample magnetic parameters and ansatz used for both textures can be found in the section Sample magnetic parameters and texture ansatz. Regarding the energy contributions, for the DW case, we included the exchange, anisotropy, and self-dipolar energy, while for the skyrmions sample, we included the previous magnetic energies together with the DMI energy. For both textures,  we approximate the self-dipolar energy as an easy-in-plane anisotropy which can be accounted for in an effective perpendicular magnetic anisotropy constant $K_{\text{eff}}=K_\text{u}-\mu_0M_\text{s}^2/2$, which is valid for a thin magnetic sample \cite{thin_film_approx_ref}.  \\

To find $l_{\text{th}}$, Eq. \eqref{eq_Fz_sample_both_textures} must be inserted into Eq. \eqref{eq:thermal_noise} and solve it numerically. For the DW case, we calculate  Eq. \eqref{eq_Fz_sample_both_textures} using Cartesian coordinates, i.e.:
\begin{align}
       F_z(\Delta) =  \pm \, \mu_0 M_\text{s} M_\text{t} R^3 L_x L_z 
\Bigg[\int_{- \infty}^{\infty} \frac{m_y(y,\Delta) (y^2 - 4d_z^2)}{(y^2+d_z^2)^{7/2}} ydy +
\int_{- \infty}^{\infty} \frac{d_z m_z(y,\Delta) (2d_z^2 - 3y^2)}{(y^2+d_z^2)^{7/2}}dy\Bigg],
    \label{eq:fz_dw}
\end{align}
while for the skyrmion case it is convenient to use cylindrical coordinates:
\begin{align}
      F_z(r_\text{s}) = \pm \, 2\pi \mu_0 M_\text{s} M_\text{t} R_\text{t}^3 
    \Bigg[\int_0^{L_z} \int_0^{r_\text{s}} \frac{\rho^2 \, d\rho \, dz \, (\rho^2 - 4(z - d_z)^2)\, m_{\rho}(\rho,r_\text{s})}{(\rho^2  + (z - d_z)^2)^{7/2}}+ \nonumber \\ 
    \int_0^{L_z} \int_0^{r_\text{s}}\frac{\rho \, d\rho \, dz \ (z - d_z)\,(3\rho^2 - 2(z-d_z)^2)\, m_z(\rho,r_\text{s})}{(\rho^2  + (z - d_z)^2)^{7/2}} \biggl], 
    \label{eq:sky_force}
\end{align}

where we have considered the tip located at a vertical distance $d_z$ respect to the sample. The details of the calculations of these equations are given in Section B of the Supplementary Material. Besides, for the selected magnetic textures, the Ritz minimizable parameters ($l_{\text{ph}}$) correspond to the domain wall width, $\Delta$, and skyrmion diameter $2 \, r_\text{s}$.\\

In Fig. \ref{fig:lth_T}, we show $l_\text{th}$ as a function of temperature for different tip radii and distances of separation $h$ for both textures. Filled markers with full lines correspond to the domain wall case, while open markers with dashed lines hold for the skyrmion sample. Note that since Eq. \eqref{eq:thermal_noise} is quadratic in the magnetic force, so that the sign choice in the tip magnetization does not produce any effect on the minimum detectable signal $l_\text{th}$ of the textures. As expected, $l_\text{th}$ increases with both the temperature and distance of separation since thermal fluctuations could induce oscillations of the cantilever and because the magnetostatic interaction decreases with the tip-sample distance, too. In general, for a given temperature, $l_\text{th}$ is larger for the skyrmion case compared to the DW sample, except for the most interacting case ($h=21$ nm) and for tip radii 15 and 20 nm, where skyrmions of length close to 5 nm could be detected at room temperature.  One can notice that for the most interacting case, there is not a valid solution for the skyrmion system for temperatures below $T  \lessapprox 38$ K and $T \lessapprox 80$ K  when $R_\text{t} = 15$ and 20 nm, respectively, since Eq. \eqref{eq:thermal_noise} does not accept valid solutions for such a set of parameters.\\
\begin{figure}[h]
\centering
\includegraphics[width=12cm]{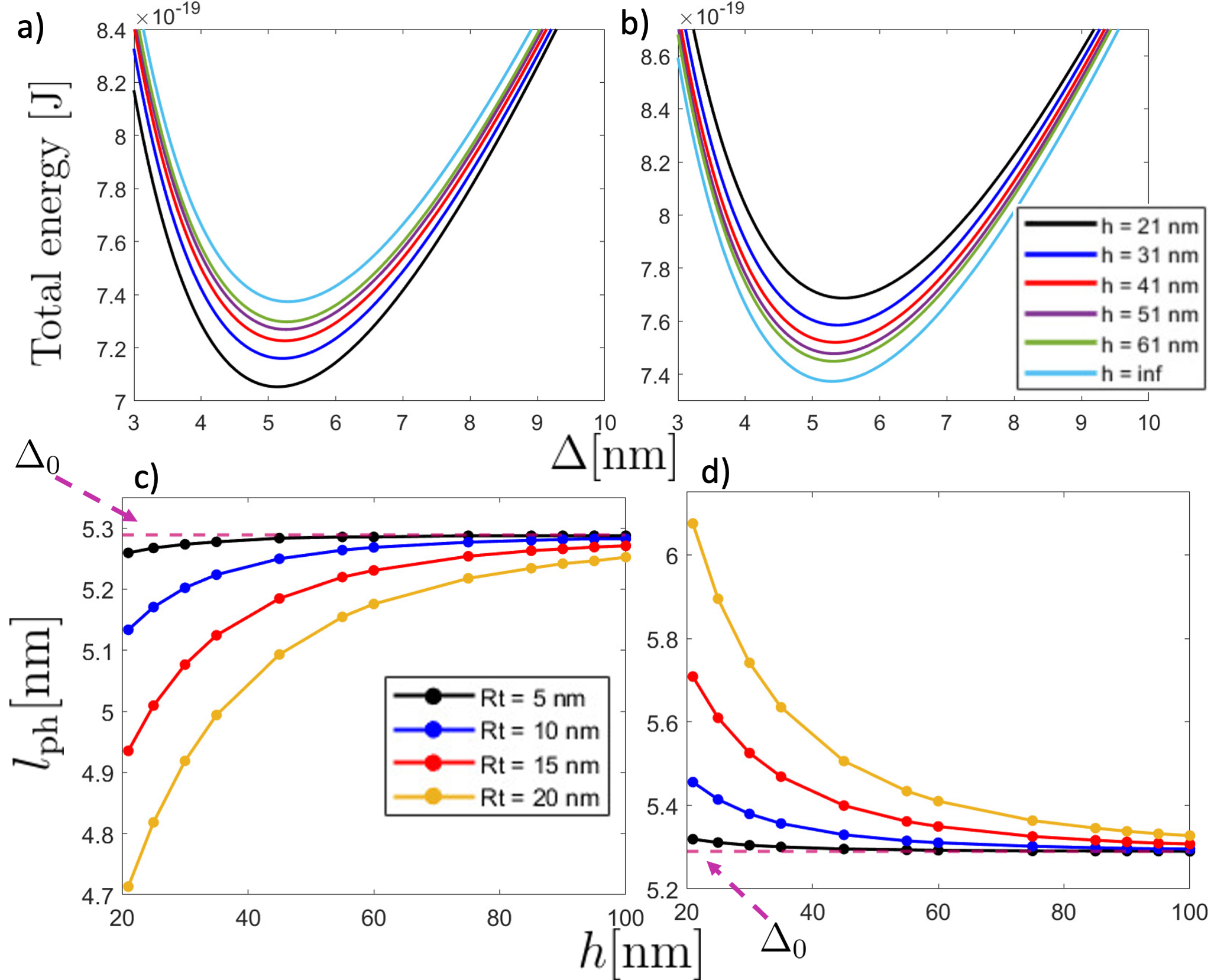}
\caption{(Upper panel) Total energy of the magnetic stripe as a function of the DW width $\Delta$ for different distances of separation with $R_\text{t}=10$ nm and considering a) $\mathbf{m}_t = -\hat{k}$ and b) $\mathbf{m}_t = +\hat{k}$. Note the difference in the horizontal scales. (Lower panel) Physical length $l_{\text{ph}}$ as a function of the distance of separation for different tip radii and considering c) $\mathbf{m}_t = -\hat{k}$ and d) $\mathbf{m}_t = +\hat{k}$. The pink dashed line stands for the specific DW width $\Delta_0$ in the absence of the magnetic tip.}
\label{fig:E_lch_dw}
\end{figure}

\begin{figure}[h!]
\centering
\includegraphics[width=12cm]{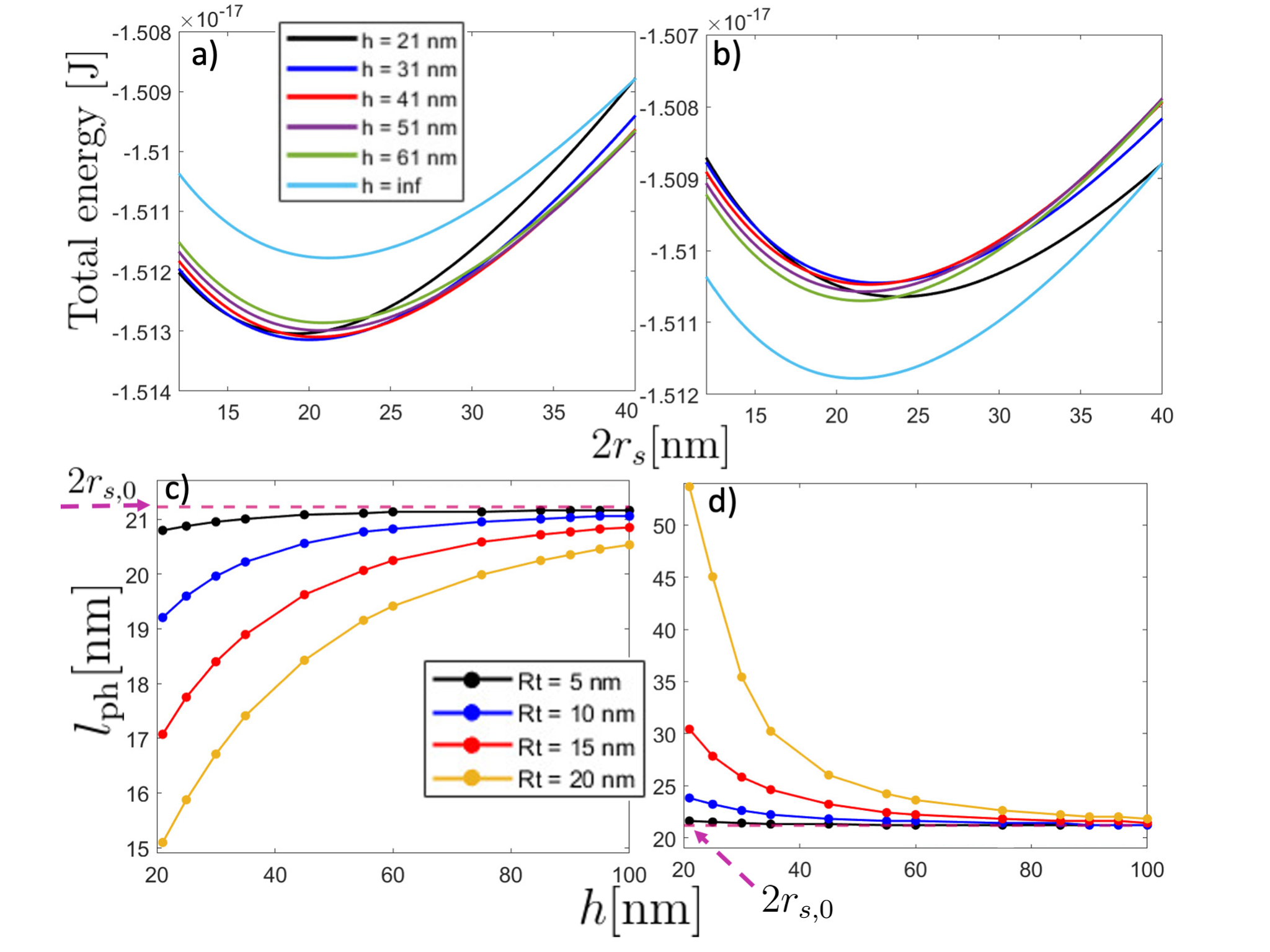} \caption{(Upper panel) Total energy of the magnetic nanodot hosting a skyrmion as a function of the skyrmion diameter $2r_\text{s}$ for different distances of separation with $R_\text{t}=10$ nm and considering a) $\mathbf{m}_t = -\hat{k}$ and b) $\mathbf{m}_t = +\hat{k}$. (Lower panel) Physical length $l_{\text{ph}}$ as a function of the distance of separation for different tip radii and considering c) $\mathbf{m}_t = -\hat{k}$ and d) $\mathbf{m}_t = +\hat{k}$. The pink dashed line pointed by the dashed arrow stands for the specific skyrmion diameter $2r_\text{s,0}=21.1$ nm in the absence of the magnetic tip.}
\label{fig:sky_E_lch_dw}
\end{figure}

Now, we calculate the physical length $l_{\text{ph}}$ that the magnetic textures acquire in the presence of the tip, which corresponds to the characteristic length that minimizes the total energy. In Figs. \ref{fig:E_lch_dw} and \ref{fig:sky_E_lch_dw} we show both the total energy and $l_{\text{ph}}$ for DW and skyrmion case, respectively. In these figures, panels a) and b) show the total energy as a function of the characteristic texture length ($\Delta$ for the DW and $2r_\text{s}$ for the Skyrmion) for different distances of separation $h$, fixing $d_y=0$ and $R_\text{t}= 10$ nm, and considering tip magnetization direction $\mathbf{M_\text{t}} = -\hat{k}$ and $ +\hat{k}$, respectively. Additionally, panels c) and d) show the physical texture length as a function of $h$ for different tip radius when $\mathbf{M_\text{t}} = -\hat{k}$ and $+\hat{k}$, respectively. As expected, $l_{\text{ph}}$ converges to the non-interacting case, i.e., $\Delta_0$ for DW and $2r_\text{s,0}$ for skyrmion, as the distance of separation becomes larger.
Regarding the DW sample, the physical DW width slightly deviates from its non-interacting value $\Delta_0 \approx 5.3$ nm when the tip points opposite to the DW magnetization direction. Furthermore, the influence of the tip on the DW size becomes more prominent as the tip radius increases and when the tip and DW magnetizations align. However, in the last case, the DW deformation is not significant as it produces small changes in DW width. To compare $l_\text{th}$ with $l_\text{ph}$, we can use Figs. \ref{fig:lth_T} and \ref{fig:E_lch_dw}c)-d). By assuming that $l_{\text{ph}}$ holds for a relatively wide range of temperatures $T\lessapprox 300$ K, we can see that the condition $l_{\text{th}}\leq l_{\text{ph}}$ never occurs. Indeed, from Fig. \ref{fig:lth_T}, one can notice that by using a tip with radius $R_\text{t} = 20$ nm with a distance of separation $h=21$ nm at room temperature, the minimal detectable signal $l_{\text{th}}$ is about 10 nm. Additionally, from Fig. \ref{fig:E_lch_dw}c), when the tip points opposite to the DW magnetization center, the physical DW width $l_{\text{ph}}$ is about 4.7 nm. In contrast, $l_{\text{ph}}$ is about 6.1 nm if the tip aligns with the DW magnetization center. Therefore, no matter the magnetization direction of the tip, the measurement will always be overestimated in this particular case.\\

On the other hand, for the skyrmion sample, the skyrmion diameter can be strongly changed respect to its non-interacting value $2r_{s,0 }\approx21.1$ nm. Note that, similar to the DW case, major deformations of skyrmion diameter occur when the tip's magnetization aligns with the skyrmion core direction ($+\hat{k}$). Here, the alignment between the magnetostatic field lines of tip and skyrmion core increases the radius of such a magnetic texture. This implies that when the magnetization of the tip is aligned to the skyrmion core magnetization and the tip approaches the sample, the skyrmion tends to increase its radius until it eventually could become a quasi-uniform magnetic state \cite{vidal2019controlling,riveros2021field} (see Fig. \ref{fig:sky_E_lch_dw}d)). Figs. \ref{fig:lth_T}, \ref{fig:sky_E_lch_dw}c), and \ref{fig:sky_E_lch_dw}d) evidence that $l_{\text{th}}$ is, at least, similar to $l_{\text{ph}}$. For instance, for the case $R_\text{t}=20$ nm evaluated at 300 K and $h=21$ nm, $l_{\text{th}}\approx 5.8$ nm (Fig. \ref{fig:lth_T}d)). For the same parameters, $l_{\text{ph}}\approx 15.2$ nm for the tip opposite to the skyrmion core (Fig. \ref{fig:sky_E_lch_dw}c)) and $l_{\text{ph}}\approx 53.7$ nm for the tip parallel to the skyrmion's core magnetization (Fig. \ref{fig:sky_E_lch_dw}d)). Therefore, in this specific case, the skyrmion size will appear about $28\%$ thinner (from its non-interacting value) when the tip is opposing the magnetization in the center of the skyrmion core and about $60\%$ larger when it is parallel to it. Unlike the domain wall texture, here we can anticipate that MFM measurements will yield reliable results because the condition $l_{\text{th}}<l_{\text{ph}}$ is fulfilled.\\

For completeness, we have compared the model for the DW stripe sample with published experimental results available in the literature. Although typical experiments are conducted under conditions beyond the model presented here, we can still make some estimations for realistic measurements. For instance, in references \cite{huo1998micromagnetic,prejbeanu2000observation,foss1996localized}, the authors show that there is a notorious difference in the domain wall width when changing the DW magnetization direction, which in turn is equivalent to changing the tip magnetization in our model. The two distinct modes are called \textit{attractive} (tip and sample magnetization pointing in the same direction) and \textit{repulsive} (tip and sample magnetization pointing in the opposite direction). Additionally,  most measurements are carried out in remanence, which our model cannot include as it implies a given magnetic history through the hysteresis loop. For DWs corresponding to metastable states, this is a standard routine to get DWs to be imaged. However, we can crudely simulate the state of remanence in the model by considering an external applied magnetic field. We first show that an external magnetic field of a few hundred mT allows us to enhance the DW width, reaching relatively similar values as the experimental reports. Indeed, for a sample with the same magnetic parameters as used in the DW case above, and considering $L_x = 500$ nm, $L_y= 1000$ nm, and $L_z=50$ nm, we get $\Delta_0\approx 48.1$ nm when applying a magnetic field of $\mathbf{B}=340 \hat{k}$ mT, as shown in Fig. \ref{fig:lph_B} of the Supplementary Material Sec. D. In such section, we also show the effect of the tip on $l_{\text{ph}}$ as a function of $h$ and the tip radius $R_\text{t}$ while the external magnetic field is applied. We have considered the two distinct configurations mentioned above, i.e., for repulsive configuration in Figs. \ref{fig:experiment}a) and \ref{fig:experiment}c), and attractive configuration Figs. \ref{fig:experiment}b) and \ref{fig:experiment}d), showing a variation in the DW width between both modes that qualitatively agrees with the experimental results in Ref. \cite{prejbeanu2000observation}. \\

\begin{figure}[h!]
\centering
\includegraphics[width=17cm]{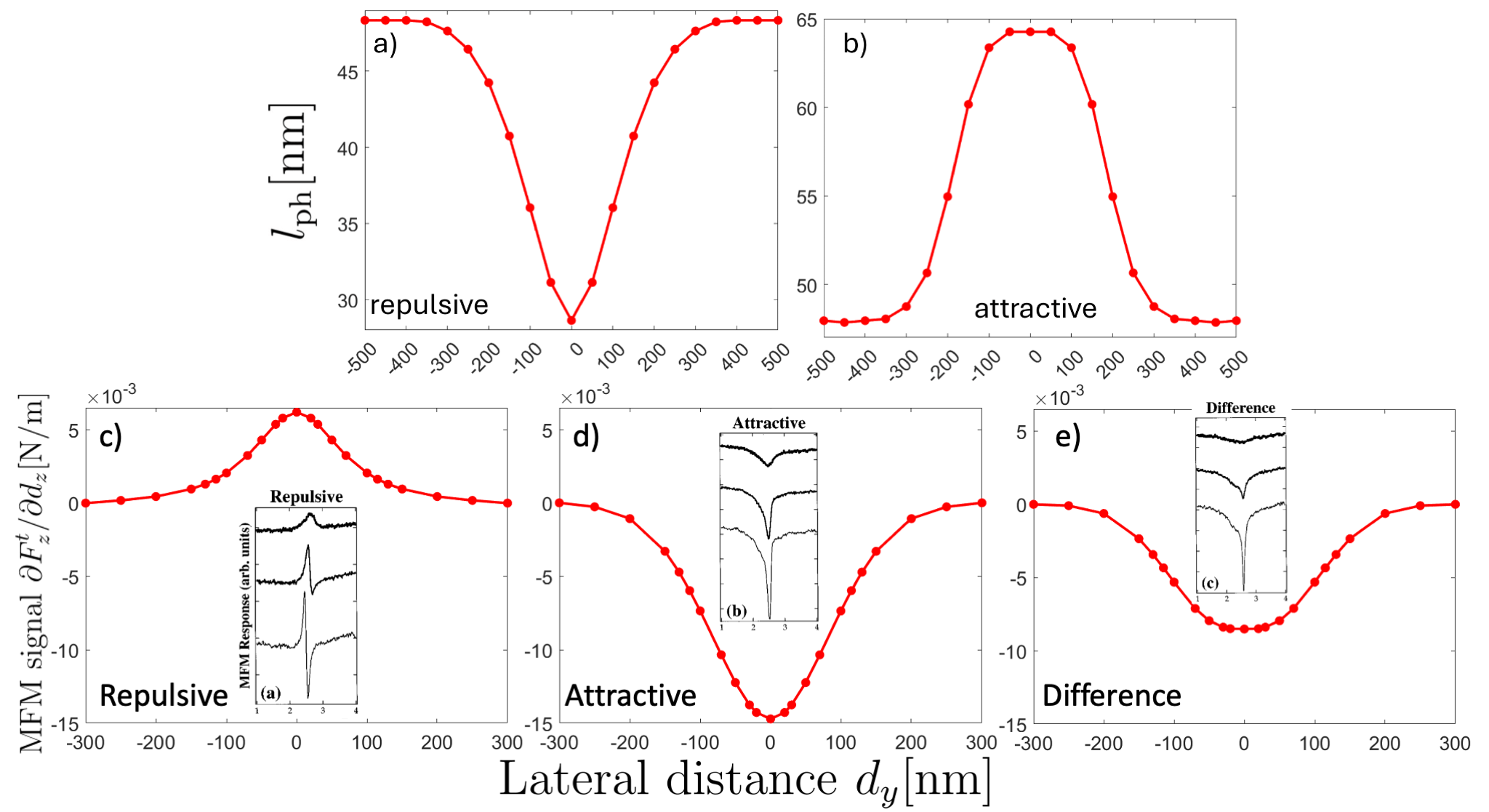} \caption{DW in stripes: Physical length $l_{\text{ph}}$ as a function of the horizontal component of the distance of separation $\mathbf{d}$ for a) $\mathbf{m}_t = -\hat{k}$ and b) $\mathbf{m}_t = +\hat{k}$, at fixed $h = 15$ nm and for $R_\text{t}=20$ nm. Force gradient over the tip as a function of the horizontal distance $d_y$ for c) repulsive and d) attractive configurations; e) shows the difference between the attractive and the inverted, repulsive profiles. The insets in Figs c), d), and e) correspond to experimental results from Fig. 3 in Ref. \cite{foss1996localized}}
\label{fig:dy_variation}
\end{figure}

Besides, we can go further by performing calculations of the MFM response, maintaining the vertical sample-tip separation fixed while varying the lateral distance $d_y$ along the nanostripe axis to estimate the effect of a horizontal scan of the tip at a fixed height over the DW. We assumed that the DW is pinned at $d_y=0$, so there are no changes in the position of the wall. Figs. \ref{fig:dy_variation}a,b) show the physical DW width, $l_{\text{ph}}$, as a function of the lateral position of the tip with respect to the DW position at a fixed height of $h = 15$ nm using a tip of radius 20 nm in the repulsive and attractive mode, respectively. Similarly, Figs. \ref{fig:dy_variation} c,d) show the MFM signal, defined as $\partial F_{\text{t},z}/\partial d_z$ \cite{foss1996localized,prejbeanu2000observation}, as a function of the lateral position $d_y$, being $F_{\text{t},z}$ the $z-$component of the magnetic force over the tip, in the repulsive and attractive modes, respectively. This force gradient is basically the same as the response obtained in experimental MFM profiles, as evidenced in Figure 3 of Ref.\cite{prejbeanu2000observation} and Figures 2 and 3 of Ref.\cite{foss1996localized}. Here, we calculated the magnetic force using Eq. \eqref{eq_Fz_sample_both_textures} replacing $l_\text{ch}$ by the corresponding $l_\text{ph}$ values and after applying an overall minus sign due to Newton's third law. As can be seen, there is a marked difference in the force gradients between the repulsive and attractive modes. In Fig. \ref{fig:dy_variation}e), we quantify such a difference by subtracting the inverted repulsive profile from the attractive one. Importantly, these are typical profiles as obtained from MFM measurements. Indeed, Figure 3 in Ref.\cite{foss1996localized} shows the same behavior for the scanning of a 180° DW. For comparison, we included explicitly these experimental results as insets in Figs. \ref{fig:dy_variation}c) - \ref{fig:dy_variation}e). It can be seen that the model results have the same behavior as the experimental results for the attractive and repulsive modes, including their differences. Nevertheless, the experimental results (insets) show greater deformations of the DW structure as long the sample-tip separation becomes smaller, evidenced by noticeable asymmetries in the MFM profile, behavior that the model can not cast as it does not include other types of deformations of the DW apart of their sizes. It is important to point out that, in addition to thermal noise, typical measurements are also limited by other noise sources attributed to the presence of lasers, amplifiers, vibrations, or tip damping. Therefore, the presence of an external magnetic field in our model intends to qualitatively mimic most of the noise source contributions.

\begin{figure}[h!]
\centering
\includegraphics[width=12cm]{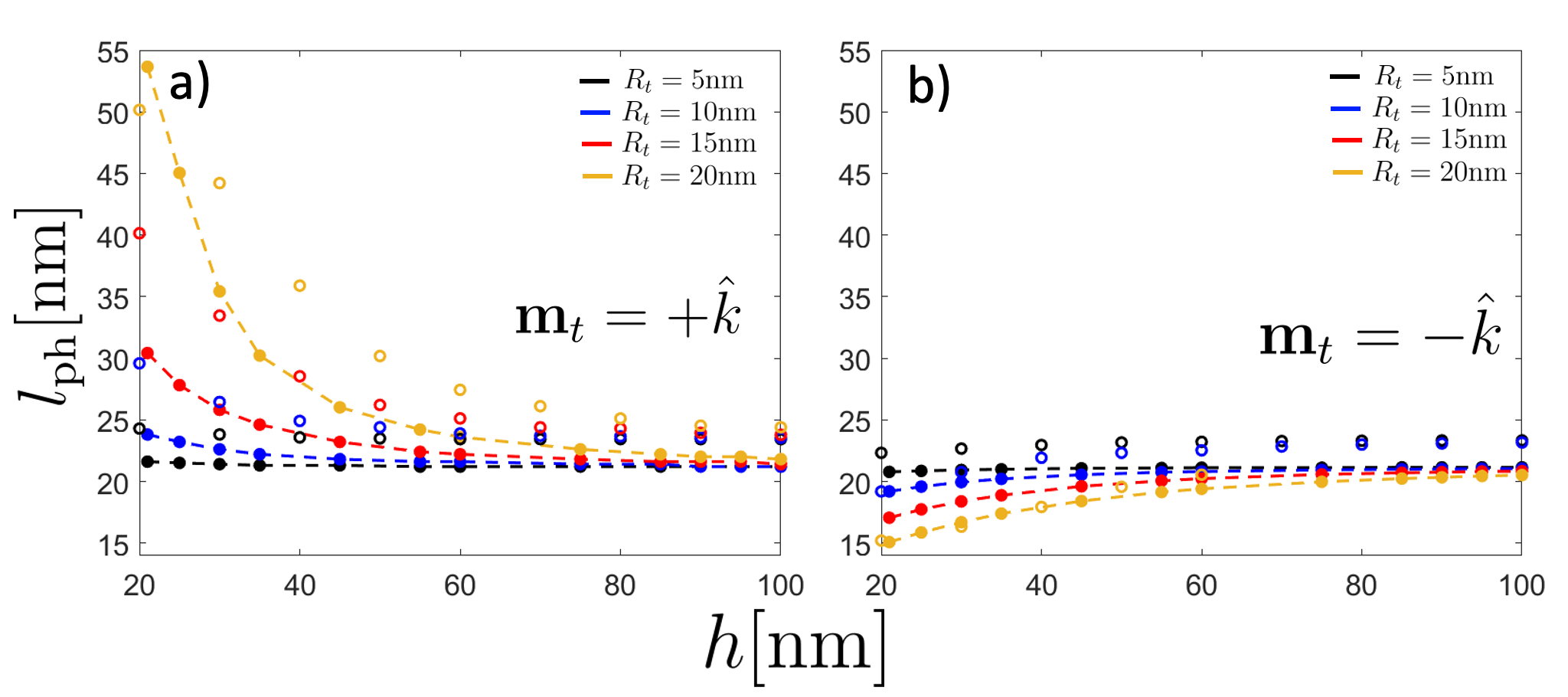} \caption{Physical skyrmion diameter $l_{\text{ph}}$ as a function of the distance of separation $h$ for different tip radius $R_\text{t}= 5$ nm (black), $R_\text{t}= 10$ nm (blue), $R_\text{t}= 15$ nm (red), and $R_\text{t}= 20$ nm (yellow). Filled symbols with dashed lines correspond to the results of the theoretical method, while open symbols correspond to micromagnetic simulations. Panel a) stands for the \textit{atractive} mode (tip magnetization pointing in the same direction of the skyrmion core), while b) stands for the \textit{repulsive} one (tip magnetization pointing opposite to the direction of the skyrmion core).}
\label{fig:simulations}
\end{figure}

Concerning the Skyrmion textures, there are some reports in which such textures are imaged and nucleated by MFM \cite{casiraghi2019individual,tan2020skyrmion,meng2019observation}.  Nevertheless, none of them pay special attention to their possible deformations. Therefore, we have compared the predicted physical skyrmion diameter in the presence of the tip with micromagnetic simulations (see details in Micromagnetic simulations section). In Fig. \ref{fig:simulations}, we show both simulations (open markers) and model results (filled markers with dashed lines) for the physical skyrmion diameter as a function of the separation distance between the tip and the sample hosting a skyrmion with positive polarity. Fig. \ref{fig:simulations}a) corresponds to the case in which both the tip and the skyrmion core point in the same direction (+$\hat{k}$), while Fig. \ref{fig:simulations}b) depicts the opposite case, in which the tip has a contrary magnetization to the skyrmion polarity. As can be seen, in both attractive and repulsive cases there is a good agreement between analytical calculations and micromagnetic simulations. 
For the non-interacting state, which can be assumed for a large $h$, the micromagnetic simulations and calculations differ by approximately 3 nm. This discrepancy can be atributed to the assumption of the self-dipolar energy as an anisotropy constant $-\mu_0M_\text{s}^2/2$, where we considered a demagnetizing factor $N_{zz}=1$. Due to the size of the nanodisk, a tiny reduction in $N_{\text{zz}}$ should be considered, which implies a small increase in the effective anisotropy constant $K_{\text{eff}}$ and, consequently, an increase of the Skyrmion diameter. On the other hand, for the attractive case (see Fig. \ref{fig:simulations}a)), the micromagnetic simulations and calculations show a more pronounced deviation of the physical length for small values of $h$. Nevertheless, in the case $R_\text{t}=20$ nm, the calculations deviates more significantly compared to the simulated case for $h<30$ nm (see Fig. \ref{fig:simulations}a)). This behavior can be attributed to the model assuming a rigid tip's magnetization, whereas in the simulations some deviations of the magnetization are expected. Such deviations reduce the magnetic charges in the spherical tip, thereby reducing the stray field. Thus, the interaction between the Skyrmion and the tip is more significant in the calculations than in the simulations.

\section*{Conclusions}
In this work, we have developed a method to theoretically characterize the deformation of an MFM measurement of the magnetic texture in the sample by considering both thermal fluctuations and the force between the magnetic tip and the sample magnetization.
The magnetization texture can be imaged when the physical characteristic length of the texture in the presence of the MFM tip ($l_{\text{ph}}$) is equal or greater than the thermal noise limited resolution ($l_{\text{th}}$) of the instrument. However, the MFM tip will still modify the texture. A measure for the influence of the tip is the difference between the observed length of the texture and the theoretical length if the tip were infinitely far away  $\vert l_{\text{ph}} - l_\text{ph} (h\rightarrow \infty) \vert$. This value depends on the tip magnetization direction through $l_{\text{ph}}$. As a result, one will overestimate the characteristic length from the MFM image when $l_\text{ph} \leq l_\text{th}$ 
whereas $l_\text{ph}$ may still be smaller than $l_\text{ph} (h\rightarrow \infty)$.

The proposed method is general and allows addressing different magnetic textures having a characteristic length whose magnetization does not vary significantly through the sample volume. Next to skyrmions these include magnetic vortices, helical states and chiral domain walls.

When applied to the case of a domain wall, our model shows qualitative agreement with published MFM measurements. By adjusting parameters such as the height of separation, the tip radius, and the inclusion of an external magnetic field, we found a good qualitative agreement between our model and experimental results in both attractive and repulsive modes. The model does not only predict the correct wall width, but also the change in wall profile for attractive or repulsive wall-tip interaction. 


To further assess our model we compared its prediction to micromagnetic simulations for a tip-skyrmion system, obtaining a maximum difference of 6 nm for the skyrmion diameter.

Our study provides an improvement of the interpretation of MFM measurements. Possible applications involving magnetic textures are thought to be functionalized or manipulated as single magnetic solitons whose magnetization structure is as unperturbed as possible, especially in the absence of magnetic fields. In this sense, it becomes essential to know the physical length of magnetic textures that could configure a necessary ingredient for future applications and, at the same time, avoid considerable deformations when exposed to MFM measurements. Hence, our findings should be taken into account when measuring characteristic lengths of different magnetic textures so that the MFM measurements are not overestimated due to the deformation induced by the presence of the magnetic tip.

\section*{Sample magnetic parameters and texture ansatz}
For the DW case, we consider a Permalloy (Py) nanostripe of size $L_x =15$ nm, $L_z=5$ nm, and $L_y\gg L_x,L_z$, with $M_\text{s} = 8.6\times 10^5$ A/m, $A_{\text{ex}}=1.3\times 10^{-11}$ J/m, and $K_\text{u}=0$ \cite{yin2006magnetocrystalline}. This choice of DW corresponds to a one-dimensional domain wall lying along the $y-$axis and whose magnetization field stands at the $y-z$ plane, as represented in Fig. \ref{fig:schematic}a), and given by $\mathbf{m}_\text{s} = m_y(y,\Delta)\hat{y} \,+ \, m_z(y,\Delta)\hat{z}$, with the Ritz function \cite{aharoni2000introduction} $m_y(y,\Delta) = \tanh \left( \frac{y - y_0}{\Delta}\right)$, where $y_0$ corresponds to the position of DW center, and $\Delta$ is the DW width. In the absence of interaction between the tip and sample, the DW width becomes $\Delta_0=\sqrt{A_\text{ex}/{|K_\text{eff}|}}$.\\

On the other hand, for the skyrmion case,  we consider a magnetic skyrmion hosted in a cylindrical Pt/Co/Pt thin nanodot as represented in Fig. \ref{fig:schematic}b). The magnetic sample is characterized by: $M_\text{s} = 580$ kA/m, $L_z=1$ nm, nanodot radius $r_\text{s}=100$ nm, $A_{\text{ex}} = 15 \times 10^{-12}$ J/m, $K_\text{u} = 0.7 \times 10^{6}$ J/m$^3$ and the interfacial Dzyaloshinskii-Moriya (DM) parameter $D = 3 \times 10^{-3}$ J/m$^2$\, \cite{tejo2018distinct}. To describe the magnetization profile of skyrmions hosted in thin films, we use a normalized magnetization vector $\mathbf{m}_\text{s}= \, m_{\rho}(\rho,r_\text{s}) \hat{\rho} + m_z(\rho,r_\text{s}) \hat{z}$, with the ansatz \cite{tejo2018distinct,vidal2019controlling,debonte1973properties} $m_z(\rho,r_\text{s}) = -P \cos[2 \arctan (f(\rho,r_\text{s}))],
    \label{eq:sky_ansatz}$, where $P =  \pm 1$ is the skyrmion polarity, which we fix equal to $P=+1$, $f(\rho,r_\text{s}) = (r_\text{s}/\rho) \exp \left( \frac{\xi}{l_{ex}} (r_\text{s} - \rho) \right)$, with $l_{ex} = \sqrt{2A_{ex}/\mu_0 M_\text{s}^2}$, $\xi^2 = C - 1$ and $C = 2K_\text{u}/\mu_0 M_\text{s}^2$. Here, $r_\text{s}$ is the skyrmion radius. Finally, for the tip and cantilever, we consider the parameters given in Refs. \cite{abelmann2005towards,tanaka2012theoretical,guimaraes2009principles}, i.e. a Cobalt spherical tip with saturation magnetization $M_\text{t}=1.44 \times 10^6 \text{A/m}$ attached to a cantilever with $Q = 3000$, $k = 2.8 \text{N/m}$, $\omega_0 = 75 \times 10^3 \, \text{Hz}$, $B = 200 \, \text{Hz}$, $A = 15$ nm.\\

\section*{Micromagnetic simulations}
We performed micromagnetic simulation for the skyrmion-tip sample using the in-house CUDA native micromagnetic solver, PETASPIN, which numerically integrates the Landau-Lifshitz-Gilbert (LLG) equation by applying the time solver scheme Adams-Bashforth \cite{giordano2012semi}. For the simulations, we used the same geometrical and magnetic parameters of the skyrmion sample in the presence of a spherical tip and to carry out the analysis, we perform two set of simulations, one to calculate the magnetostatic field due to the MFM spherical tip, and another one to calculate the effect of the magnetostatic field due to the tip on the skyrmion size.
For the former, we consider the spherical tip placed on top of the sample and centered respect to disk center, such that we are able to compute the magnetostatic field outside the tip for a distance ranging from 0 to 100 nm. The tip is considered with a uniform magnetization along either the positive or negative out-of-plane direction z-axis, and we use a cubic discretization cell of $1\times 1 \times 1$ nm$^3$. For the latter, we perform simulations of the disk, with initial state as a Néel skyrmion  placed in the middle of the disk. 
From the first set of simulations, we extract the magnetostatic field at the specific distance from the tip and we apply this non-uniform field as an external field to the skyrmion-based disk. The skyrmion radius was calculated as the radius of the area enclosed within the core skyrmion region.

\section*{Acknowledgements}
This work was supported by ANID Fondecyt Iniciación 11220046, Fondecyt Regular 1210607and 1201491, and CIP2022036. Besides we aknowledge the project number 101070287 - SWAN-on-chip - HORIZON-CL4-2021-DIGITAL-EMERGING-01, the project PRIN 2020LWPKH7 "The Italian factory of micromagnetic modelling and spintronics" and the project PRIN20222N9A73 "SKYrmion-based magnetic tunnel junction to design a temperature SENSor-SkySens", funded by the Italian Ministry of University and Research (MUR) and by the PETASPIN Association (www.petaspin.com). DR, RT and MC acknowledge the support from the project PE0000021, "Network 4 Energy Sustainable Transition - NEST", funded by the European Union - NextGenerationEU, under the National Recovery and Resilience Plan (NRRP), Mission 4 Component 2 Investment 1.3 - Call for Tender No. 1561 dated 11.10.2022 of the Italian MUR (CUP C93C22005230007). DR also acknowledges the support of the project D.M. 10/08/2021 n. 1062 (PON Ricerca e Innovazione), funded by the Italian MUR.


\bibliography{Wt}
\clearpage
\section*{Supplementary Material}
In this supplementary material, we show major details of the calculation of the magnetostatic field generated by the spherical tip, interaction energy, coordinates transformation, total micromagnetic energy, the numerical evaluation of $l_{\text{th}}$ and the effect of an additional external magnetic field on $l_\text{ph}$ for the DW sample in presence of the tip.
\appendix
\section{Magnetostatic field generated by an uniformly magnetized spherical tip}
The picture of the system coordinates form sample and tip center is depicted in Fig. \ref{fig:Ap1}. 
\renewcommand{\thefigure}{A\arabic{figure}}

\setcounter{figure}{0}
\begin{figure}[h]
\centering
\includegraphics[width=8.6cm]{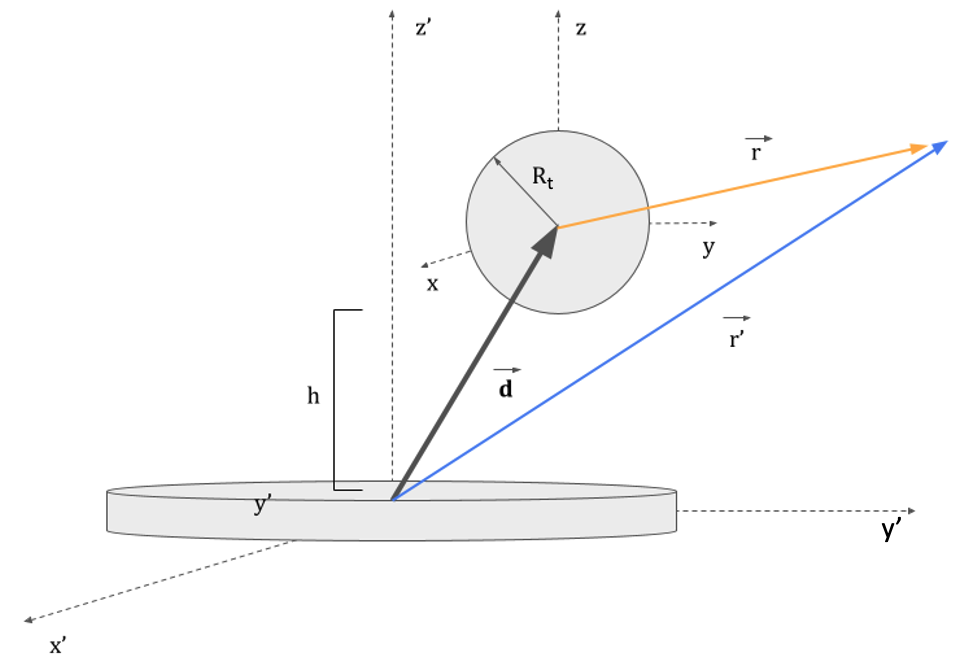}
\caption{Schematic representation of the system studied. The frame $x'y'z'$ is fixed at the sample (which corresponds to a nanodisk or a nanostrip), while the frame $xyz$ is fixed to the spherical tip}
\label{fig:Ap1}
\end{figure}

Since we have chosen to carry out the integral involved in Eq. 2 on the sample volume $V_s$, we must to express both $\mathbf{m}_\text{s}$ and $\mathbf{h}_\text{t}$ in a coordinate system solidarity to the sample. We start by computing the magnetostatic field generated by the spherical tip as seen in a reference frame fixed at its origin. In this case, $\mathbf{h}_\text{t}=-\nabla U(\mathbf{r})$, with the magnetostatic potential

\begin{align}
   U(\mathbf{r}) = \frac{1}{4\pi}\left[-\int_{V'} \frac{\mathbf{\nabla'}\cdot\mathbf{M}(\mathbf{r}')}{|\mathbf{r}-\mathbf{r}'|} \,dv' + \int_{S'}\frac{\mathbf{M}(\mathbf{r}') \cdot \hat{n}'}{|\mathbf{r}-\mathbf{r}'|}\,ds'\right],
\end{align}
where $V'$ ($S'$) stands for the sphere volume (surface). Considering the homogeneous magnetization of tip pointing to the $\pm z$ direction, the volume integral vanishes. On the other hand, the spherical symmetry can be exploited to expand the Green function $|\mathbf{r}-\mathbf{r}'|^{-1}$ as

\begin{align}
\frac{1}{|\mathbf{r}-\mathbf{r}'|} = \sum_{l=1}^{\infty}\sum_{m=-l}^{l} \frac{4\pi}{2l + 1} \frac{r_<^l}{r_>^{l+1}} \; Y_{l,m}(\theta,\phi)Y_{l,-m}(\theta',\phi'),
\end{align}
where $Y_{l,m}(\theta,\phi)$ are the spherical harmonic functions, which are given by
\begin{align}
Y_{l,m}(\theta,\phi) = \sqrt{\frac{2l+1}{4\pi}\frac{(l-m)!}{(l+m)!}}\;P_l^m(\cos\theta)e^{im\phi},
\end{align}
with $P_l^m(\cos\theta)$ the Legendre Polynomials. Since in our case $\hat{n}' = \hat{r}'$, the magnetostatic potential can be written as follows

\begin{align}
\label{eqn:u3}
    U(\mathbf{r}) = \frac{M_\text{t}}{4\pi}\sum_{l=1}^{\infty}\sum_{m=-l}^{l} \frac{r_<^l}{r_>^{l+1}} \;P_l^m(\cos\theta)e^{im\phi}\int_{S'} \cos\theta'\;P_l^{-m}(\cos\theta')e^{-im\phi'} \,ds'    
\end{align}
By using the identity $ 2\pi\; \delta_{0,m} =\int_{0}^{2\pi} e^{-im\phi'} \,d\phi'$,
and exploiting the orthogonality of Legendre Polynomials, we get the following expression for the magnetostatic potential

\begin{align}
        U(\mathbf{r}) =  \frac{M_\text{t}}{3}\frac{r_<}{r_>^2} \;R_\text{t}^2\cos\theta
\end{align}

As described above, we need to find an expression for $\mathbf{h}_\text{t}$ on the sample reference system. To do that, we start by calculating the magnetostatic potential in the region outer the sphere as seen on its own reference system. In this region $R_\text{t} < r$. Thus, the magnetostatic potencial is given by $U(\mathbf{r}) =  \frac{M_\text{t} R^3}{3}\;\frac{\cos\theta}{r^2}$, from which we can formally get the magnetostatic field

\begin{align}
     \mathbf{H}_{t}(\mathbf{r}) = \frac{M_\text{t}R_\text{t}^3}{3r^3} [2\cos\theta \hat r + \sin\theta \hat\theta],
\end{align}
that corresponds to Eq. (5) of the main text. 
This expression can be written in cartesian coordinates

\begin{align}
\label{eqn:h}
    \mathbf{H}_\text{t}(\mathbf{r}) = \frac{M_\text{t}R_\text{t}^3}{(x^2 + y^2 + z^2)^{5/2}} \left[ xz\, \hat{x} +yz \, \hat{y} + \frac{1}{3}(2z^2 -x^2 -y^2)\, \hat{z} \right]
\end{align}

By using the following transformation between sample and tip reference frames (see Fig. \ref{fig:Ap1}), $x' = x$ ,  $y' = y + d_y$,  $z' = z + d_z$, where $d_y$ and $d_z$ are the components of the $\mathbf{d}$ vector mentioned in the main text, we can find the expression for $\mathbf{H}_{t}$ in the sample reference system. The expression for the stray field in cartesian and cylindrical coordenates is given by

\begin{align}
\begin{split}
    \mathbf{h}_\text{t}(\mathbf{r'}) = & \frac{M_\text{t} R_\text{t}^3}{(x'\,^2 + (y' - d_y)^2 + (z' - d_z)^2)^{5/2}} \\& \left[ x'(z' - d_z)\, \hat{x'} \, + (y' - d_y)(z' - d_z)\, \hat{y'}\,' + \frac{1}{3}(2(z' - d_z)^2 - x'\,^2 - (y' - d_y)^2) \, \hat z' \right] 
\label{eq:campoh_cartesianas}
\end{split}
\end{align}

\begin{align}
\begin{split}
    \mathbf{h}_\text{t}(\mathbf{r'}) = & \frac{M_\text{t} R_\text{t}^3}{(\rho'\,^2\cos^2 \phi ' + (\rho'\sin \phi ' - d_y)^2 + (z' - d_z)^2)^{5/2}}  \biggl((\rho ' - d_y\sin \phi')(z' - d_z)\, \hat \rho' \\& - d_y \cos \phi ' (z' - d_z) \, \hat \phi'
    + \frac{1}{3}(2(z' - d_z)^2 - \rho'\,^2 \cos ^2 \phi ' - (\rho' \sin \phi ' - d_y) ^2)\, \, \hat z' \biggl)
    \label{eq:campoh_cilindricas}
\end{split}
\end{align}
It can be easily noted that if one want to considerate the opposite direction $+z$ for the tip magnetization, these results will only change in a global sign.

\section{Total micromagnetic energy of the sample and magnetic force}
The total micromagnetic energy comprises the exchange, magnetostatic, effective anisotropy (including dipolar energy), Dzyaloshiinski-Moriya, and interaction energy. The latter is obtained by integrating the dot point of Eq. (A11) or (A12)  with the sample magnetization. The rest of energies can be calculated as 

\begin{align}
    E_\text{ex} = A_\text{ex} \int  \sum_i (  \mathbf{\nabla} m_i )^2 \, dV_\text{s},
\label{exchange}
\end{align}
where $A_\text{ex}$ is the stiffness constant, and $m_i$ is the ith-component of the sample magnetization with $i = x, y, z$. The magnetic anisotropy energy comprises the contribution from an easy axis (proportional to $K_\text{u}$) and an easy plane, which emerges from a magnetostatic approximation proportional to $M_\text{s}^2$
\begin{align}
    E_\text{ani} = - K_{\text{eff}} \int m_z^2 \, dV_\text{s},
\label{eq:anisotropy}
\end{align}
where $K_{\text{eff}}= K_\text{u} - \mu_0 \, M_\text{s}^2 / 2$ is the effective anisotropy constant, with $K_\text{u}$ the uniaxial anisotropy constant, and the second term corresponds to an approximation to the dipolar interaction of the sample. Finally, the interfacial DMI energy is given by 
\begin{align}
E_\text{DM} = - D \int [m_z \, \vec{\nabla} \cdot \vec{m} - (\vec{m} \cdot \vec{\nabla})\, m_z] \, dV_\text{s},
\label{eq:e_d-m}
\end{align}
where $D$ is the DMI parameter.

\subsection{Domain Walls}
For the case of DWs, it is convenient to use the expression for $\mathbf{h}_\text{t}$ written in cartesian coordinates. By considering the following ansatz for the DW magnetization field
\begin{align}
    m_y(y,\Delta) = \tanh \biggl( \frac{y - y_0}{\Delta} \biggl) ,
\end{align}
we arrive to the interaction energy
\begin{align}
    E_\text{int} = \mp \, \frac{\mu_0 M_\text{s} M_\text{t} L_x L_z R_\text{t}^3}{3}\int_{- \infty}^{\infty} \frac{\sqrt{1 - \tanh^2 \biggl( \frac{y - y_0}{\Delta} \biggl)}(2d_z^2 - (y - d_y)^2) - 3d_z \tanh \biggl( \frac{y - y_0}{\Delta} \biggl)(y-d_y)}{((y - d_y)^2 + d_z^2)^{5/2}} \,dy ,
\end{align}
where the negative sign will be considered if $\mathbf{M}_t$ is pointing in the $+z$ direction, and the positive sign will be considered if $\mathbf{M}_t$ is pointing in the opposite $-z$ direction. Next, we can now calculate the $z$ component of the magnetic force on the sample and its derivative respect to the $d_z$ distance
\begin{align}
       F_z(\Delta) =  \pm \, \mu_0 M_\text{s} M_\text{t} R^3 L_x L_z 
\Bigg[\int_{- \infty}^{\infty} \frac{m_y(y) (y-d_y) ((y - d_y)^2 - 4d_z^2)}{((y-d_y)^2+d_z^2)^{7/2}}dy +
\int_{- \infty}^{\infty} \frac{d_z m_z(y) (2d_z^2 - 3(y - d_y)^2)}{((y-d_y)^2+d_z^2)^{7/2}}dy\Bigg],
    \label{eq:fz_dw}
\end{align}

\begin{align}
    \begin{split}
      F_z =  \pm \, \mu_0 M_\text{s} M_\text{t} R_\text{t}^3 L_x L_z \int_{- \infty}^{\infty} \frac{\tanh \biggl( \frac{y - y_0}{\Delta} \biggl) (y-d_y) ((y - d_y)^2 - 4d_z^2) +d_z\, \sqrt{1 - \tanh^2 \biggl( \frac{y - y_0}{\Delta} \biggl)} (2d_z^2 - 3(y - d_y)^2)}{((y-d_y)^2+d_z^2)^{7/2}}
      \, dy
    \end{split}
    \label{eq:fz_dw_appen}
\end{align}

\begin{align}
\begin{split}
    \frac{\partial F_z}{\partial d_z} = \pm \, \mu_0 M_\text{s} M_\text{t} R_\text{t}^3 L_x L_z \int_{- \infty}^{\infty} & \biggl[ \frac{\tanh \biggl( \frac{y - y_0}{\Delta} \biggl) (20d_z^3(y-d_y) - 15d_z(y-d_y)^3)}{((y-d_y)^2 + d_z^2)^{9/2}} \\& + \frac{\sqrt{1 - \tanh^2 \biggl( \frac{y - y_0}{\Delta} \biggl)} (24d_z^2 (y-d_y)^2 - 8d_z^4 - 3(y-d_y)^4)}{((y-d_y)^2 + d_z^2)^{9/2}}  \biggl] \, dy
    \label{eq:dfz_dw_appen}
\end{split}
\end{align}

By using the ansatz (B4), the exchange and magnetic anisotropy energies reads

\begin{align}
\label{eq:ex_dw}
    E_{ex} = \frac{2A_\text{ex} \, L_x L_z}{\Delta},
\end{align}

\begin{align}
\label{eq:eani_dw}
    E_\text{ani} = - 2K_{\text{eff}} \, L_x L_z \Delta.
\end{align}

\subsection{Skyrmions}
For this magnetic texture we must proceed analogously to the previous case. In this case we use the following ansatz
\begin{align}
        m_z(\rho,r_\text{s}) = -P \cos[2 \arctan (f(\rho,r_\text{s}))] = -P \cos[2 \arctan ((r_\text{s}/\rho) \, \text{exp}(\xi(r_\text{s} - \rho)/l_{\text{ex}})],
\end{align}
whose parameters are introduced in the main text. Due to the symmetry of the nanodisk, it is convenient to use the expression for $\mathbf{h}_\text{t}$ in cylindrical coordinates. The interaction energy between the tip and sample can be written as follows
\begin{align}
\begin{split}
    E_{\text{int}} = \mp \, \frac{\mu_0 M_\text{s} M_\text{t} R_\text{t}^3}{3} \int_0^{L_z} \int_0^{2\pi} \int_0^{r_\text{s}} & \frac{\rho}{(\rho^2 \cos^2 \phi + (\rho \sin \phi - d_y)^2 + (z - d_z)^2)^{5/2}} \\& \biggl[ 3\, \sqrt{1 - P^2 \cos^2[2 \arctan ((r_\text{s}/\rho) \, \text{exp}(\xi(r_\text{s} - \rho)/l_{\text{ex}})]}(\rho - d_y\sin \phi)(z - d_z)  \\& -P \cos[2 \arctan ((r_\text{s}/\rho) \, \text{exp}(\xi(r_\text{s} - \rho)/l_{\text{ex}})] \, (2(z - d_z)^2 - \rho^2 \cos^2 \phi \\& - (\rho \sin\phi - d_y)^2)  \biggl] \, d\rho \, d\phi \, dz.
\end{split}
\end{align}
Therefore, the $z$ component of the magnetic force on the sample and its derivative are

\begin{align}
      F_z(r_\text{s}) = \pm \, 2\pi \mu_0 M_\text{s} M_\text{t} R_\text{t}^3 
    \Bigg[\int_0^{L_z} \int_0^{r_\text{s}} \frac{\rho^2 \, d\rho \, dz \, (\rho^2 - 4(z - d_z)^2)\, m_{\rho}(\rho,r_\text{s})}{(\rho^2  + (z - d_z)^2)^{7/2}}+ \nonumber \\ 
    \int_0^{L_z} \int_0^{r_\text{s}}\frac{\rho \, d\rho \, dz \ (z - d_z)\,(3\rho^2 - 2(z-d_z)^2)\, m_z(\rho,r_\text{s})}{(\rho^2  + (z - d_z)^2)^{7/2}} \biggl], 
    \label{eq:sky_force}
\end{align}

\begin{align}
\begin{split}
    F_z = \pm \, 2\pi \mu_0 M_\text{s} M_\text{t} R_\text{t}^3 \int_0^{L_z} \int_0^{r_\text{s}} & \frac{\rho \, d\rho \, dz}{(\rho^2 \cos^2 \phi + (\rho \sin \phi - d_y)^2 + (z - d_z)^2)^{7/2}} \biggl[ (\rho  - d_y \sin\phi) (\rho^2 \cos^2 \phi + (\rho \sin \phi - d_y)^2 \\& - 4(z - d_z)^2)\sqrt{1 - P^2 \cos^2[2 \arctan ((r_\text{s}/\rho) \, \text{exp}(\xi(r_\text{s} - \rho)/l_{\text{ex}})]}  \\& - (z - d_z)\,(3\rho^2 \cos^2\phi + 3(\rho\sin\phi - d_y)^2 \\& - 2(z-d_z)^2) P \cos[2 \arctan ((r_\text{s}/\rho) \, \text{exp}(\xi(r_\text{s} - \rho)/l_{\text{ex}})] \biggl], 
\end{split}
\end{align}

\begin{align}
\begin{split}
\frac{\partial F_z}{\partial d_z} = \pm \, 2\pi \mu_0 M_\text{s} M_\text{t} R_\text{t}^3 \int_0^{L_z} \int_0^{r_\text{s}}  & \frac{d\rho \, dz}{(\rho^2 \cos^2 \phi + (\rho \sin \phi - d_y)^2 + (z - d_z)^2)^{9/2}} \\& \biggl[ 5(\rho - d_y \sin\phi)(z - d_z)(3\rho^2 \cos^2\phi + 3(\rho\sin\phi - d_y)^2 \\& - 4(z-d_z)^2) \sqrt{1 - P^2 \cos^2[2 \arctan ((r_\text{s}/\rho) \, \text{exp}(\xi(r_\text{s} - \rho)/l_{\text{ex}})]}  - (24\rho^2 \cos^2 \phi (z-d_z)^2 \\& + 24(\rho\sin\phi - d_y)^2 (z-d_z)^2- 8(z-d_z)^4 -3\rho^4\cos^4\phi - 3(\rho\sin\phi - d_y)^4 \\& - 6\rho^2 \cos^2\phi(\rho\sin\phi - d_y)^2) \, P \cos[2 \arctan ((r_\text{s}/\rho) \, \text{exp}(\xi(r_\text{s} - \rho)/l_{\text{ex}})] \biggl]
\end{split}
\end{align}

Finally, the expressions for exchange energy, anisotropy, and DMI can be now obtained. In cylindrical coordinates, the exchange energy reads

\begin{align}
\begin{split}
    E_{\text{ex}} = 2\pi L_z A_{\text{ex}} \int_0^{r_\text{s}} \biggl[\biggl( & \frac{1}{\rho} (1 - P^2 \cos^2[2 \arctan ((r_\text{s}/\rho) \, \text{exp}(\xi(r_\text{s} - \rho)/l_{\text{ex}})]) \biggl) \\& + \frac{4 P^2 r^2_s \,\text{exp}^2(\frac{\xi}{l_{\text{ex}}}(r_\text{s} - \rho)) (\frac{1}{\rho} + \frac{\xi}{l_{\text{ex}}})^2 \sin^2 (2 \arctan (\frac{r_\text{s}}{\rho}\text{exp}(\frac{\xi}{l_{\text{ex}}}(r_\text{s} - \rho))))}{\rho \, (1 + (\frac{r_\text{s}}{\rho} \text{exp}(\frac{\xi}{l_{\text{ex}}} (r_\text{s} - \rho)))^2)^2 \, (1 - P^2 \cos^2[2 \arctan ((r_\text{s}/\rho) \, \text{exp}(\xi(r_\text{s} - \rho)/l_{\text{ex}})])} \, \biggr] d\rho,
\end{split}
\end{align}
Next, the effective anisotropy energy can be easily calculated
\begin{align}
    E_{\text{ani}} = -2\pi L_z K_{\text{eff}}\int_0^{r_\text{s}} \rho \, P^2 \cos^2[2 \arctan ((r_\text{s}/\rho) \, \text{exp}(\xi(r_\text{s} - \rho)/l_{\text{ex}})] \, d\rho
\end{align}

Finally, the DMI is

\begin{align}
\begin{split}
    E_{\text{DM}} = -2\pi L_z D \int_0^{r_\text{s}} & \biggl[ \frac{2 P^2 \cos^2[2 \arctan ((r_\text{s}/\rho) \, \text{exp}(\xi(r_\text{s} - \rho)/l_{\text{ex}})] - 1}{\sqrt{1 - P^2 \cos^2[2 \arctan ((r_\text{s}/\rho) \, \text{exp}(\xi(r_\text{s} - \rho)/l_{\text{ex}})]}}  \\& \frac{2Pr_\text{s} \, \text{exp}(\frac{\xi}{l_{\text{ex}}}(r_\text{s} - \rho))(\frac{1}{\rho} + \frac{\xi}{l_{\text{ex}}}) \, \sin(2\arctan(\frac{r_\text{s}}{\rho}\text{exp}(\frac{\xi}{l_\text{ex}}(r_\text{s} - \rho))))}{1 + (\frac{r_\text{s}}{\rho} \, \text{exp}(\frac{\xi}{l_\text{ex}} (r_\text{s} - \rho)))^2} \\& + P \cos[2 \arctan ((r_\text{s}/\rho) \, \text{exp}(\xi(r_\text{s} - \rho)/l_{\text{ex}})] \sqrt{1 - P^2 \cos^2[2 \arctan ((r_\text{s}/\rho) \, \text{exp}(\xi(r_\text{s} - \rho)/l_{\text{ex}})]}  \biggl] \, d\rho.
\end{split}
\end{align}
\section{Calculation of $l_{\text{th}}$}
Here we show the numerical solution of Eq. \eqref{eq:thermal_noise} by matching both sides of it. For the DW case, we depict the numerical evaluation of Eq. \eqref{eq:thermal_noise} as a function of $\Delta$ by matching the left-hand side (solid curves) with the right-hand one (dashed curves) for selected distances of separation $h$ and tip radii $R_\text{t}$. According to Eq. \eqref{eq:thermal_noise}, the corresponding $l_\text{th}$ can be extracted from the value of $\Delta$ where both solid and dashed curves intersect. It is important to point out that, due to the shape of solid curves, there are more than one $\Delta$ at which the curves match. However, as previously done in Refs. \cite{abelmann2005towards,tanaka2012theoretical,porthun1998magnetic,porthun1998optimization,saito2003high,schonenberger1990understanding,hug1998quantitative} , we only consider those points that satisfy the fact that the $l_{\text{th}}$ must increase with the temperature. 

\begin{figure}[h]
\centering
\includegraphics[width=14cm]{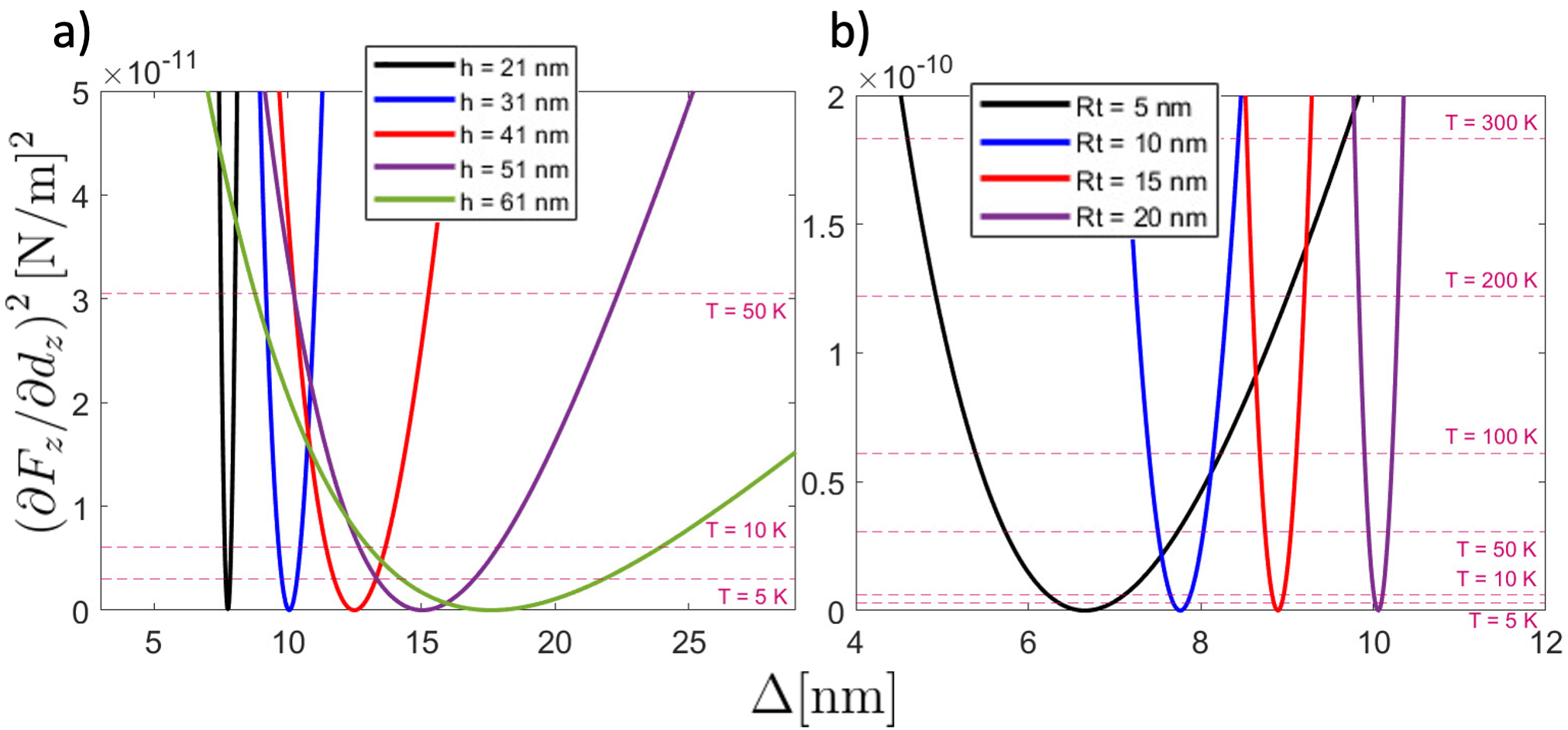}
\caption{Numerical evaluation of Eq. \eqref{eq:thermal_noise}. The solid curves correspond to the left-hand side of Eq. \eqref{eq:thermal_noise} as a function of the DW width $\Delta$ for a) $R_\text{t}=10$ nm and different lengths of separation $h$; and b) $h=21$ nm for different tip radius $R_\text{t}$. The pink dashed line corresponds to the right-hand side of Eq. \eqref{eq:thermal_noise} for different temperatures. The value of $\Delta$ at which the solid and dashed lines intersect corresponds to $l_{\text{th}}$.}
\label{fig:dfz_delta}
\end{figure}

\begin{figure}[h]
\centering
\includegraphics[width=14cm]{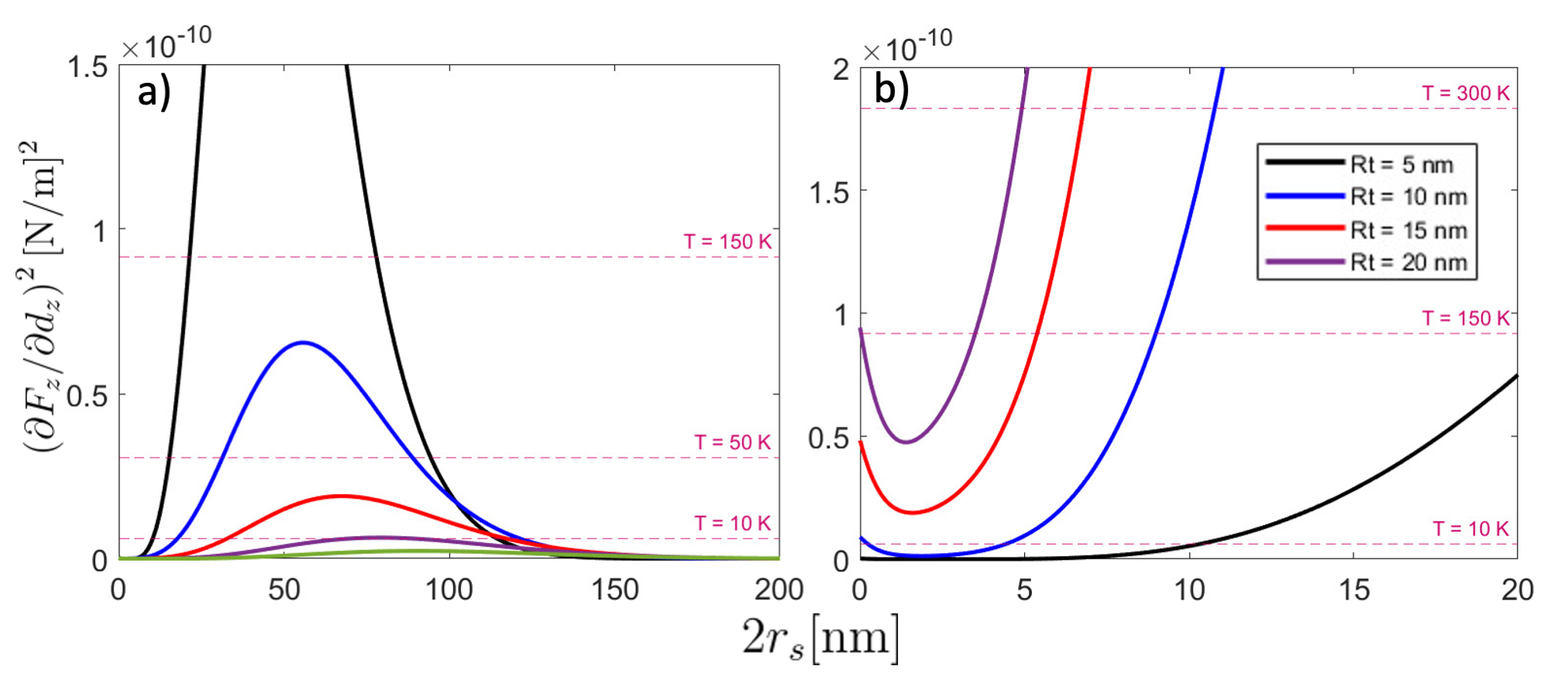}
\caption{Numerical evaluation of Eq. \eqref{eq:thermal_noise} for the skyrmion case. The solid curves correspond to the left-hand side of Eq. \eqref{eq:thermal_noise} as a function of the skyrmion diameter $2r_\text{s}$ for a) $R_\text{t}=5$ nm and different lengths of separation $h$; and b) $h=21$ nm for different tip radius $R_\text{t}$. The pink dashed line corresponds to the right-hand side of Eq. \eqref{eq:thermal_noise} at distinct temperatures. The value of $2r_\text{s}$ at which the solid and dashed lines intersect corresponds to $l_{\text{th}}$.}
\label{fig:dfz_2rs}
\end{figure}
The skyrmion case is shown in Fig. \ref{fig:dfz_2rs}. Fig. \ref{fig:dfz_2rs}a) corresponds to the solution for different heights of separation $h$ and a fixed $R_\text{t} = 5$ nm, while in Fig. \ref{fig:dfz_2rs}a) the height is fixed to $h=21$ nm, and the solution is plotted for different tip radii $R_\text{t}$. Similar to the DW case, we consider those values of $2r_\text{s}=l_{\text{th}}$ that satisfy the fact of becoming larger as the temperature increases.
\newpage
\section{Variation of the DW width with the external magnetic field}
\begin{figure}[h]
\centering
\includegraphics[width=9cm]{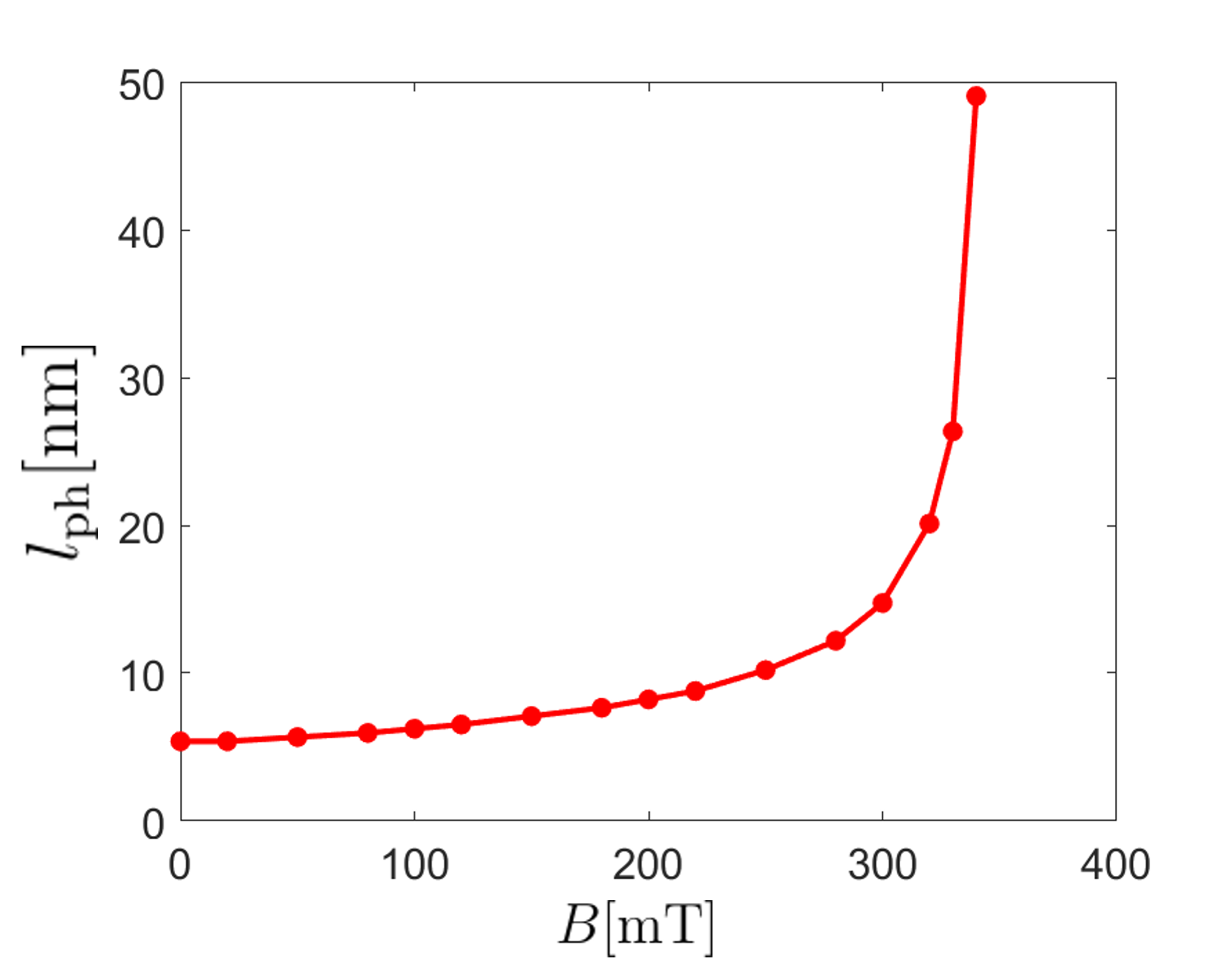}
\caption{Physical length $l_{\text{ph}}$ as a function of an external magnetic field.}
\label{fig:lph_B}
\end{figure}

Here we show the variation of the DW width that minimized the total magnetic energy as a function of an external magnetic field applied in the $z-$direction. As can be seen from Fig. \ref{fig:lph_B}, the DW width enhances because it points in the same direction as the magnetic field. This enhancement allows us to simulate the remanence state in real experiments. 
\begin{figure}[h]
\centering
\includegraphics[width=13cm]{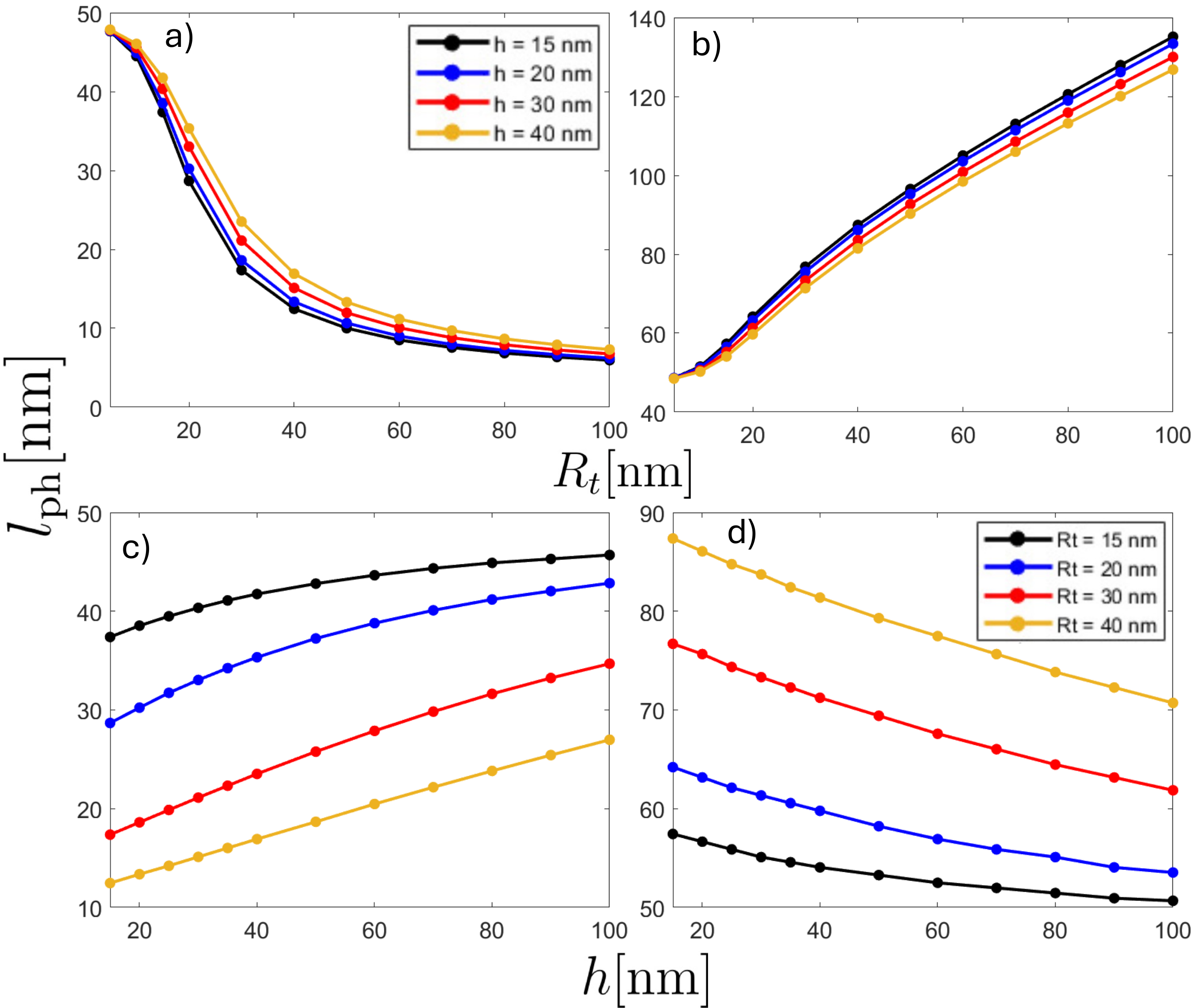} \caption{(Upper panel) Physical length $l_{\text{ph}}$ of the DW as a function of the tip radius for different height of separation for a) $\mathbf{m}_t = -\hat{k}$ and b) $\mathbf{m}_t = +\hat{k}$. (Lower panel) Physical length $l_{\text{ph}}$ of the DW as a function of the distance of separation for different tip radii and considering tip magnetization direction c) $\mathbf{m}_t = -\hat{k}$ and d) $\mathbf{m}_t = +\hat{k}$. All cases were calculated under the application of a magnetic field of $\mathbf{B}= 340 \text{ mT} \hat{k}$.}
\label{fig:experiment}
\end{figure}

The effect of tip on the DW width is shown in Fig. \ref{fig:experiment}, where we show $l_{\text{ph}}$, as a function of the height of separation under an applied field of 340 mT for repulsive configuration  \ref{fig:experiment}a,c), and attractive configuration Figs.  \ref{fig:experiment}b,d). In this case there exist a larger DW deformations according to the relative configuration between the sample and the tip's magnetization. For example, when the tip points parallel to the DW magnetization (attractive configuration), and a tip radius of 40 nm is employed in measurement at $h=40$ nm, the resulting DW width is about 80 nm, representing an increment of about $60\%$. While, when the tip points opposite to the DW magnetization (repulsive configuration) at same tip radius and height, the resulting DW width is about 15 nm, representing a decrement of about $70\%$. This behavior is also observed in real measurements. Indeed, Ref. \cite{prejbeanu2000observation} show an increment of approximately $60\%$ of DW width in attractive mode respect to repulsive repulsive one in samples with similar sizes as presented here. This variation qualitatively agrees with the results shown above, and as we claim here, the deformation degree can vary according to the tip geometry and the height of separation at which the measurement is carried out.
\end{document}